\documentclass[]{spie} 
\pdfoutput=1
\usepackage[]{graphicx}
\title{Universality of transition temperatures in families of copper-oxide superconductors: interlayer tunneling redux} 
\author{Angela Kopp and Sudip Chakravarty
\skiplinehalf
Department of Physics and Astronomy, University of California Los Angeles, Los Angeles, California 90095-1547, USA}
\authorinfo{Further author information: (Send correspondence to S.C.)\\S.C.: E-mail: sudip@physics.ucla.edu}
\begin{document} 
\maketitle 

\begin{abstract}
We revisit the interlayer tunneling theory of high temperature superconductors and formulate it as a mechanism by which the striking systematics of the transition temperature within a given homologous series can be understood. We pay attention not only to the enhancement of pairing, as was originally suggested, but also to the role of competing   order parameters that tend to suppress superconductivity, and to the charge imbalance between inequivalent
 outer and inner CuO$_2$ planes in a unit cell. Calculations based on a generalized Ginzburg-Landau theory yield results that bear robust and remarkable resemblance to experimental observations.
 \end{abstract}

\section{INTRODUCTION}
\label{sect:intro}  

From the point of view of a theorist  one of the hardest quantities to calculate is the transition temperature of any phase transition, because it is non-universal and depends on many details. Nonetheless, the fascination for room temperature superconductivity has been a major driving force behind superconductivity research. Thus one might wonder if a theorist has any role whatsoever to play in this affair and one may be justifiably skeptical. A theorist often faces two classes of questions: (1) Why does the substitution of La by Y in a cuprate superconductor raise the transition temperature from 30K (LBCO) to 90 K (YBCO)? (2) What separates conventional superconductors from high-$\mathrm{T_{c}}$ cuprate superconductors? The answer to the first question may be far more difficult than the second. You might also ask why changing D (deoxyribo) to R (ribo) in DNA (deoxyribonucleic acid) changes its properties dramatically? DNA can replicate but RNA cannot. The sensitivity to chemical elements is not special to high-$\mathrm{T_{c}}$ superconductors. The $\mathrm{T_{c}}$ changes from 16 K to 9.6 K going from NbN to NbTi. This is very typical of emergent phenomena; the properties of a complex system are more than the sum of its parts. Although in common parlance we often like to label a material by its chemical composition, the actual properties are vastly diverse even with respect to small changes. Diversity and specificity are common themes in emergent matter. The second question, one hopes, is relatively more tractable. There may be a fundamental shift in the mechanism---from phonon mediated superconductivity to something else. And this shift a theorist should be able to spot, but not without some clues. These clues come not only from detailed electronic, magnetic and structural data, but also from hints of patterns. In the past twenty years, much effort has been invested in unravelling details, but little in the patterns or the  universalities. The aim of this work is to analyze an universal property of high-$\mathrm{T_{c}}$ superconductors.

\begin{figure}[htb]
\begin{minipage}[t]{6.5in}
\includegraphics[width=3in]{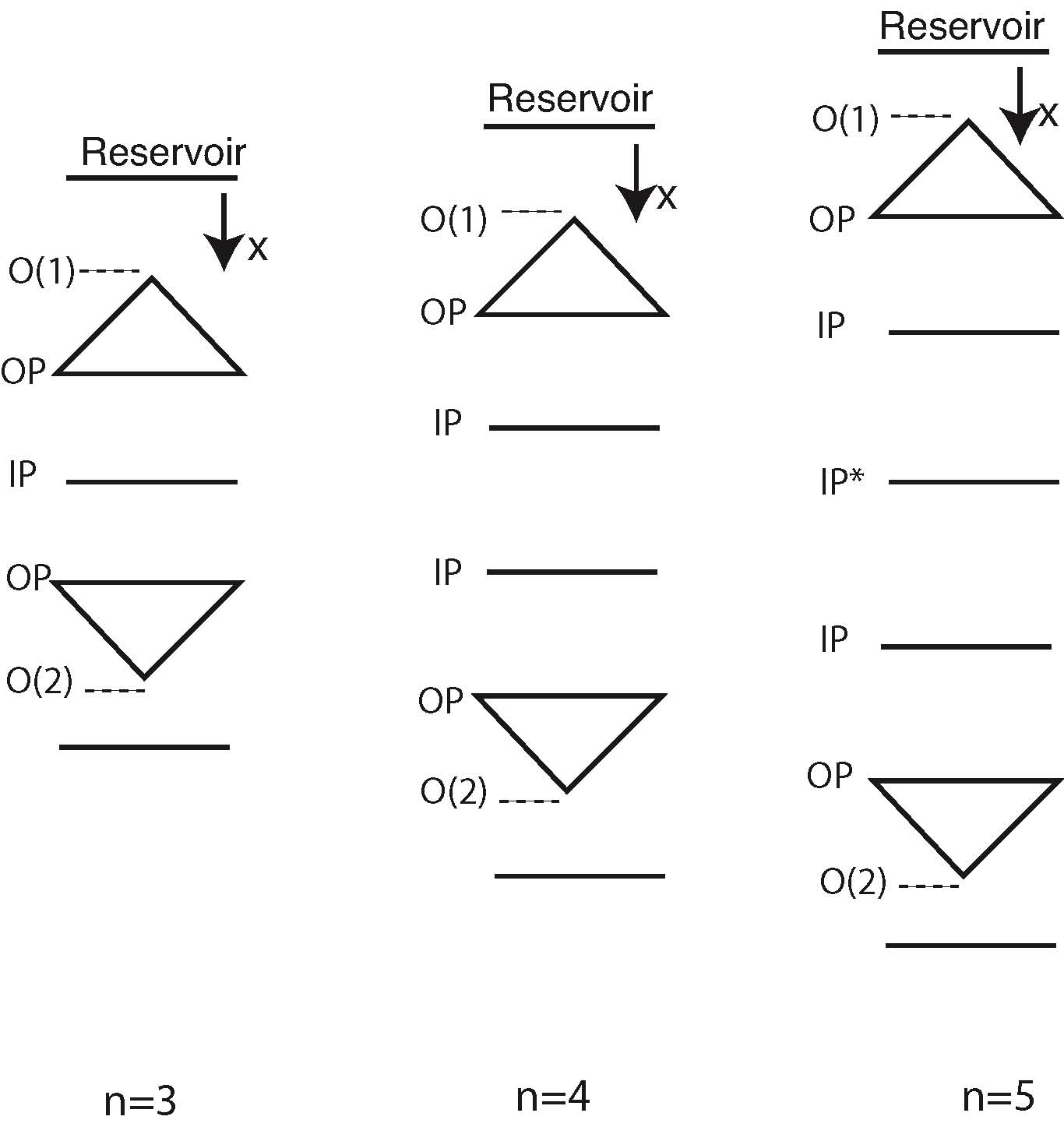} \hfill \includegraphics[width=3in]{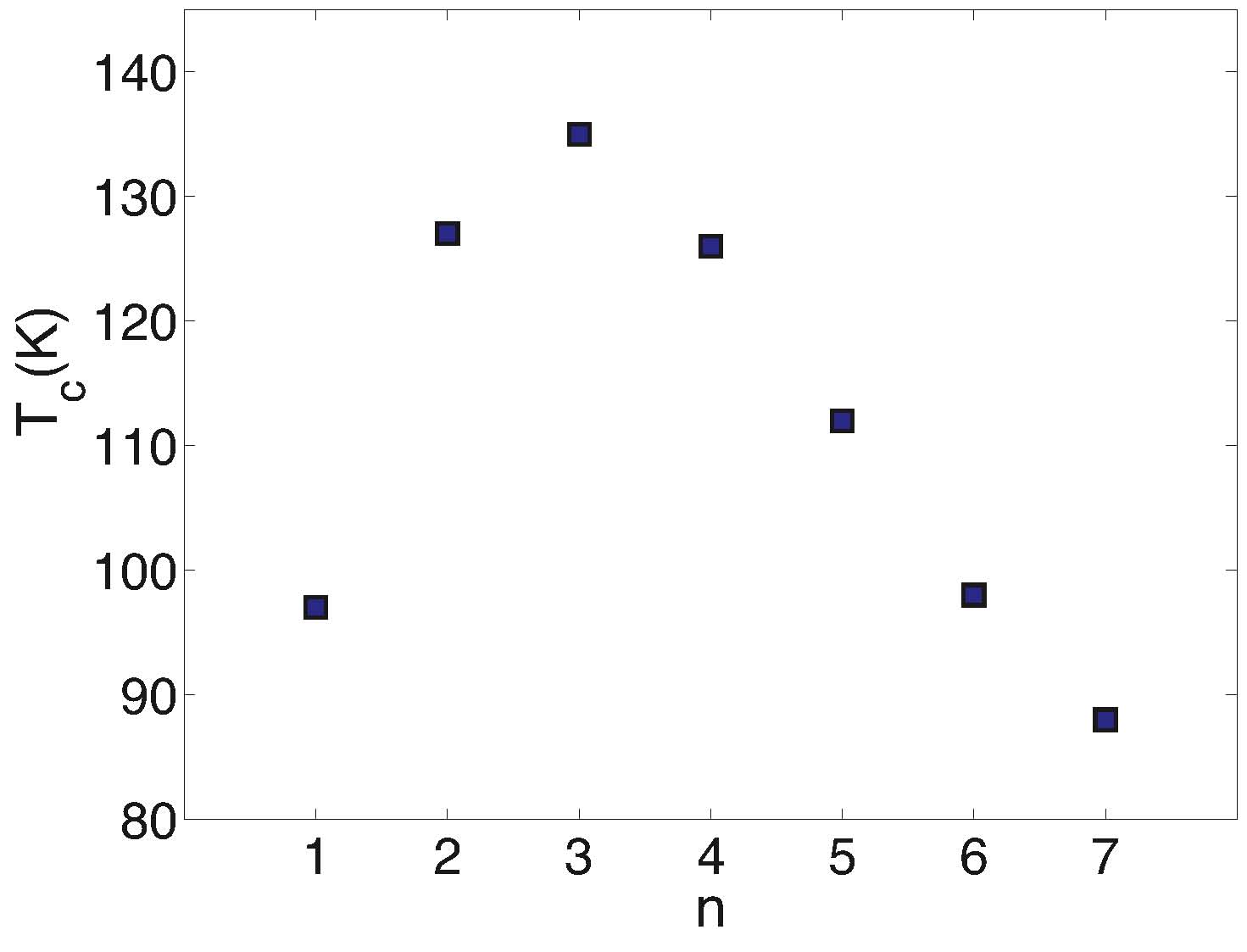}
\end{minipage}
\caption[multilayer]{\label{fig:multilayer}Left: a schematic sketch of a homologous series. OP stands for the  $\mathrm{CuO_{2}}$-planes next to  the apical oxygen and IP stands for the inner planes. Note that starting at $n=5$ one can in principle distinguish IP from IP$^{*}$. In actual experiments the doping of IP and IP$^{*}$ are very close. The nominal doping level is denoted by $x$. }
\caption[Tc]{\label{fig:Tc}Right: optimal transition temperatures within a homologous series,
HgBa$_2$Ca$_{n-1}$Cu$_n$O$_{2n+2+\delta}$, as a function of $n$, adapted from Kuzemskaya {\em et al.}~\cite{Kuzemskaya00} }
\end{figure}

The phase diagram of high temperature superconductors shows two striking universalities.  First, as a function of doping, $x$, $\mathrm{T_{c}}$  is a dome-shaped curve rising at about 5\%  and dropping to zero at 30\%. The second universal feature is its dependence on the number of $\mathrm{CuO_{2}}$-layers, $n$, within a homologous series (see Fig.~\ref{fig:multilayer}), as shown in Fig.~\ref{fig:Tc}. 
The increase in $\mathrm{T_{c}}$  in going from $n=1$ to $n=3$, about 35\%, may seem very modest  but in absolute numbers it is about 35 K. Considering that the highest transition temperature of the conventional superconductor $\mathrm{Nb_{3}Ge}$ is about 23 K, this enhancement is spectacular. It is something that is hard to ignore, although most theories of high-$\mathrm{T_{c}}$ superconductors tend to ignore this fact and little experimental attention is paid to multilayer cuprates. There are exceptions, however \cite{Kivelson02,Leggett99}. The interlayer tunneling theory was motivated by precisely this fact \cite{Chakravarty93}. 

Why is there a superconducting dome in the first place? A moment's reflection will convince one that this is quite unusual from the perspective of the BCS mechanism: any arbitrary attractive interaction leads to superconductivity.  The key to this puzzle, we believe, is competing order. A weakly interacting Fermi liquid has very little ordering tendency, simply because most perturbations around it tend to die out. Once we leave the regime of validity of a Fermi liquid, we need to deal with many length scales denoted by their characteristic coupling constants. It is almost a truism that {\em in a strongly correlated system any symmetry that can be broken must be broken}. If we ascribe superconductivity to strong correlation effects, a corollary is that a competing broken symmetry state may rear its head. Understanding competing order is a key to understanding high-$\mathrm{T_{c}}$. 

A recent paper by Chakravarty, Kee, and V\"{o}lker \cite{Chakravarty04} traces the second universality to the combination of enhancement due to interlayer tunneling and  charge imbalance between the layers as $n$ increases. If we denote the $\mathrm{CuO_{2}}$-planes closest to the apical oxygens as outer planes (OP) and the others as the inner planes (IP),  the charge carriers in a nominally optimally doped sample prefer to reside on the OP's, leaving the OP's overdoped and the IP's underdoped.\cite{Distasio90,Haines92}   The aim of the present work is to follow up on these results in greater depth. 

This paper is not a typical conference paper because it reports a substantial amount of original research not found elsewhere. In Sec. 2, we set the stage by a discussion of the conventional Josephson coupling energy. Sec. 3 is the central section in which an effective Hamiltonian coupling the planes of a high-$\mathrm{T_{c}}$ superconductor is developed. This derivation has not been reported previously. Using the results of Sec. 3, we develop the Ginzburg-Landau functional in Sec. 4 and apply it to the topic announced in the title. Some of the results discussed here have also not been previously reported. In Sec. 5 we present our final thoughts and in the Appendix we discuss a technical derivation.

\section{The Fly in the ointment}

In this section, we shall learn an important lesson from the Josephson coupling energy between two identical conventional BCS superconductors. Assuming a momentum non-conserving tunneling matrix element $T_{\bf p \bf q}$ between two identical $s$-wave superconductors, the coupling energy $\Delta E$ is\footnote{The derivation of this formula\cite{Tinkham96} in the limit that the coherence length is very long compared to the separation between  the superconductors and close to the transition temperature of the bulk follows from the Ginzburg-Landau theory.}
\begin{equation}
\Delta E = E_{J} \left(1 - \cos \phi \right),
\label{eq:deltae}
\end{equation}
where at zero temperature $E_J$ is given by
\begin{equation}
E_{J} =\frac{1}{2}\frac{R_{Q}}{R_{n}}|\Delta| \propto |\psi|, \; \; \frac{1}{R_{n}}=4\pi e^{2}N^{2}(0)\langle |T_{\bf p \bf q}|^{2} \rangle, \; \; R_{Q}=\frac{h}{4e^{2}}\approx 6.5 \mathrm{K\Omega},
\end{equation}
where $N^{2}(0)$ is the square of the density of states at the Fermi energy, and $\langle |T_{\bf p \bf q}|^{2}\rangle$ is a suitable average of the tunneling matrix element over the momentum space close to the Fermi energy. The two important points are: (1) the lowest energy is zero when the phase difference, $\phi$, between the left and the right superconductors is zero, and (2) the coupling energy is non-analytic in the order parameter, $\psi$, or the superconducting gap, $\Delta$. Although the phase-dependent term lowers the energy, the  first term cancels this lowering identically. 
Interlayer tunneling cannot enhance superconductivity in a conventional layered superconductor by directly lowering the system energy! What it can do is suppress phase fluctuations, which can indirectly enhance $\mathrm{T_{c}}$.

\begin{figure}[htb]
\begin{center}
\includegraphics[width=2.5in]{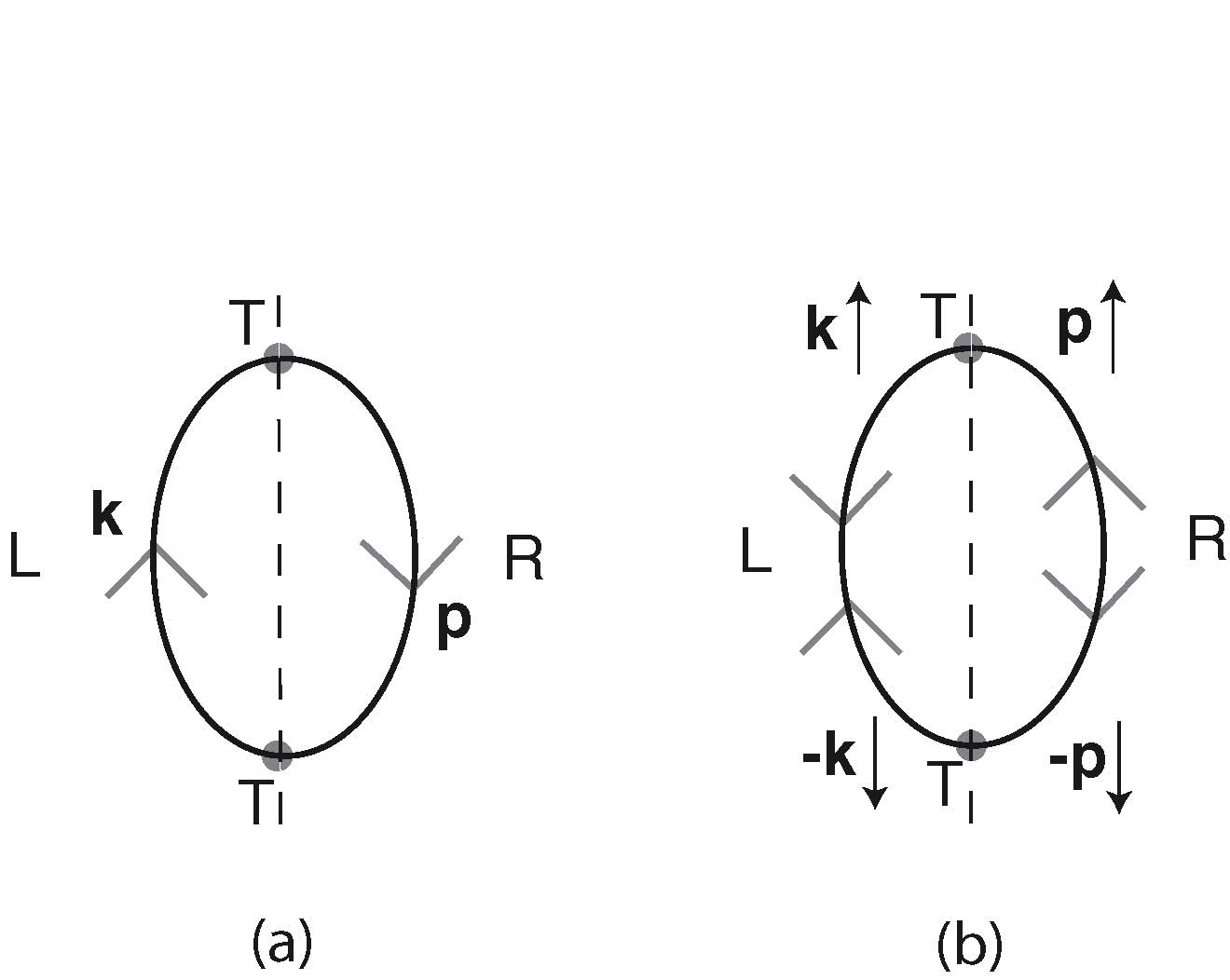} 
\end{center}
\caption[Feynman]{\label{fig:Feynman}Feynman diagrams for calculating the Josephson coupling energy between two identical superconductors on the left (L) and the right (R). The diagram (a) corresponds to single particle hops from left to right and back, and the diagram (b) corresponds to pair hopping. Both are virtual processes. The normal component of the superconducting Green function has a single arrow and the anomalous component has double arrow. The momentum non-conserving tunneling vertices are denoted by T. }
\end{figure}

There is a nice microscopic interpretation as to why this is the case, as shown in the Feynman diagrams in Fig.~\ref{fig:Feynman}.\footnote{We thank D. J. Scalapino for an illuminating discussion of this point.} 
These diagrams can be calculated in a textbook fashion\cite{Abrikosov75} using continuum superconducting Green functions without electron-electron interaction.  The momentum integrations are replaced by energy integrations by making use of density of states; the result is Eq.~\ref{eq:deltae}. When $\phi=0$, the virtual tunneling of the pairs and the single electrons both lower the energy of the system. However, because of the superconducting gap, the single electron tunneling does not lower the energy of the system as much as it would in the normal state, leading to a net positive term that exactly cancels the contribution from pair tunneling. It is possible that this exact cancellation 
is a result of the simplifications mentioned above. 
In any case, regardless of whether or not the cancellation is exact, there is an important lesson here, 
which is that virtual single electron processes compete with the pair hopping processes.
It is this {\it competition} that will be crucial in what we shall have to say later. 
The second remarkable point is the non-analyticity of the result. There is no analytic expansion of the ground state energy at $T=0$ in terms of the order parameter. It is also easy to show that close to the transition temperature $E_{J}\propto |\psi|^{2}$, and the analyticity is restored.\cite{Chakravarty94} Thus, there is no analog of the Lawrence-Doniach\cite{Lawrence71} model at $T=0$, while there is one close to $\mathrm T_{c}$. This non-analyticity is a very special feature of a Fermi liquid.

\section{Effective Hamiltonian}
In this section we shall derive an instantaneous 
effective Hamiltonian that describes coupled layered cuprate superconductors. 
This will be carried out perturbatively in the tunneling Hamiltonian, $H_T$, while, 
in principle, arbitrary electron-electron interaction will be allowed. 
This is really the key: the strong interactions are not treated perturbatively. 
Unlike the Hamiltonian for the conventional Josephson effect between two macroscopic superconductors, $H_T$ will be assumed to conserve momentum. There are no reasons for it to be momentum non-conserving, unless we wish to describe the effects of impurities.
Surprisingly, this perturbative derivation is well controlled in a number of circumstances. It is important to stress, however, that  for a normal Fermi liquid this perturbation theory breaks down completely due to vanishing energy denominators. 

The derivation of this effective low-energy Hamiltonian is important for arriving at the correct Ginzburg-Landau functional at $T=0$, which we then use to discuss the universality of the transition temperature. In principle, it is possible to use this effective Hamiltonian in a more microscopic language, but owing to the complexity of the problem we avoid this path.

\subsection{A pedagogic example}
In the previous section we have noted that in the conventional Josephson effect, the lowering of the energy due to the uncertainty principle as manifested in both single electron tunneling and pair tunneling completely cancel each other when the phase difference, $\phi = 0$. This is {\it not} a generic result, as we illustrate with an analogous simple problem.

Consider the two-site ($i=1,2$) Hubbard model for which the Hamiltonian is
\begin{equation}
H_{\mathrm 2-site}=\varepsilon_{0}\sum_{i\sigma}c_{i\sigma}^{\dagger}c_{i\sigma}- t\sum_{\sigma}\left(c_{1,\sigma}^{\dagger}c_{2,\sigma}+\mathrm{h. c.}\right)+U\sum_{i} n_{i\uparrow}n_{i\downarrow},
\end{equation}   
where  $c_{i,\sigma}^{\dagger}$ creates an electron at site $i$ with spin $\sigma$, and $n_{i\uparrow}$ is the number operator for an up spin at site $i$. When the on-site Coulomb interaction, $U=0$, the problem cannot be treated in perturbation theory in $t$ because of degeneracy (vanishing energy denominators), but an exact solution shows that the symmetric state is lower in energy than the antisymmetric state by $2t$. In contrast, when $U/t\to \infty$ and the system is half full (one electron per site), one should include the on-site interaction in the  unperturbed Hamiltonian  and treat the tunneling term as the perturbation. So, to zeroth order, the state of lowest energy corresponds to each site being occupied by one electron and is of energy $2\varepsilon_{0}$. The excited states correspond to one site being empty and the other doubly occupied.  The leading contribution to the ground state energy is second order in $t$, because the first order matrix element vanishes. The 4-fold degeneracy of the unperturbed ground state can in principle be lifted in second order degenerate perturbation theory:
\begin{equation}
E_{0}=2\varepsilon_{0}-\sum_{m\ne 0}\frac{|\langle 0|H_{T}|m\rangle|^{2}}{U} = 2\varepsilon_{0}-\frac{2|\langle 0|H_{T}^{2}|0\rangle|^{2}}{U}=\left\langle 0\left| H_{0}+\frac{4t^{2}}{U}\left( \mathbf{S}_{1}\cdot \mathbf{S}_{2}-\frac{1}{4}\right)\right| 0\right \rangle.
\end{equation}
Note that we have made use of the completeness relation and the fact that the tunneling Hamiltonian has no diagonal matrix element. As far as the low energy, 4-fold degenerate subspace is concerned, we have identified the effective Hamiltonian:
\begin{equation}
H_{\mathrm{eff}}=H_{0}+\frac{4t^{2}}{U} \left(\mathbf{S}_{1}\cdot \mathbf{S}_{2}-\frac{1}{4}\right),
\end{equation}
where $\mathbf{S}_{i}$ is the spin operator at site $i$.
The  terms generated in second order perturbation theory are analogous to the terms generated in the Josephson case. In the Heisenberg term an electron makes a virtual transition to the neighboring site with opposite spin and the electron on that site replaces the original electron (exchange process).  In the second term an electron makes a virtual transition and bounces back (direct process). In the singlet state, the energy is 
\begin{equation}
E_{s}=2\varepsilon_{0}-\frac{4t^{2}}{U},
\end{equation}
but in the triplet state
\begin{equation}
E_{t}=2\varepsilon_{0}.
\end{equation}
because of the cancellation of the two processes.
There are two important lessons: (1) whether or not the effective Hamiltonian lowers the energy depends on the state, and (2) the lowering is non-perturbative in $U$. In this particular case of a finite system, we could have equally well treated $U$ as a perturbation. After resummation we will arrive at the same answer. For infinitely many degrees of freedom this is far from being guaranteed.

\subsection{Similarity transformation}

Let the total  Hamiltonian be denoted
by
\begin{equation}
H=H_1+H_2+H_{T},
\end{equation}
where $H_1$ and $H_2$ are arbitrary interacting Hamiltonians describing the 
layers and $H_T$ 
the tunneling Hamiltonian coupling the layers. 
Using a similarity transformation, it is 
possible to eliminate the  tunneling Hamiltonian in favor of
processes involving two particles, or a particle and a hole, and,
of course, an infinite series of such  higher order processes. A recent
development in this direction is  infinitesimal similarity 
transformations, pioneered by Wegner \cite{Wegner98} and Wilson and Glazek \cite{Glazek93}. This technique is yet to be applied to the present problem. Here we shall 
consider only a one-step similarity transformation:
\begin{equation}
 H_{\mathrm{eff}}=e^{-S}He^S=H+[H,S]+{1\over
2}[[H,S],S]+\cdots.
\end{equation}
Here $S$ is an anti-Hermitian operator.
If we choose 
\begin{equation}
H_T+[H_{1}+H_{2},S]=0,
\end{equation}
then
\begin{equation}
H_{\mathrm{eff}}=H_{1}+H_{2}+{1\over 2}[H_T,S]+\cdots
\end{equation}
In the representation in which $H_{1}+H_{2}$ is diagonal,
\begin{equation}
\langle n|S|m\rangle=\frac{\langle n|H_T|m\rangle}{E_m-E_n},
\end{equation}
provided the energy denominator does not vanish.  As mentioned earlier, we consider a momentum conserving tunneling Hamiltonian, where
\begin{equation}
H_{T}=\sum_{{\bf k},\sigma}t_{\perp}({\bf k})(c_{{\bf
k},\sigma}^{\dagger (1)}c_{{\bf
k},\sigma}^{(2)}+{\rm h. c.}).
\end{equation}
The superscripts refer to the layer index and $\mathbf{k}=(k_{x},k_{y})$ is the in-plane crystal momentum.  For high-$\mathrm{T_{c}}$ superconductors, the tunneling matrix elements between adjacent layers are \cite{Chakravarty93,Andersen95,Pavarini01}:
\begin{eqnarray}
t_{\perp}({\bf k})&=&\frac{t_{\perp}}{4}\left[\cos (k_xa)-\cos (k_ya)\right]^2: \mathrm{tetragonal}
\label{eq:tmatrix1}
\\
t_{\perp}({\bf k})&=&\frac{t_{\perp}}{4}\left[\cos (k_xa)-\cos (k_ya)\right]^2\cos(\frac{1}{2}k_{x}a)\cos(\frac{1}{2}k_{y}a): \textrm{body centered tetragonal (bct)}
\label{eq:tmatrix2}
\end{eqnarray}
For simplicity, we shall consider only the tetragonal case. The bct case requires similar but more careful consideration of the coupling constants.
For two coupled gapless Fermi liquids, a simple analysis shows
that a momentum  conserving tunneling matrix element leads to divergences in  the  series resulting from the similarity transformation because of the
degeneracy of the single particle states between the layers. 

In general, if we denote the degenerate subspace of $(H_{1}+H_{2})$ by $|0\rangle \equiv |0\rangle_{1}\otimes  |0\rangle_{2}$, we get
\begin{equation}
\langle 0|{H_{\mathrm{eff}}}|0\rangle=\langle0|H_{1}+H_{2}|0 \rangle-\sum_{n\ne 0}\frac{|\langle
0|\sum_{{\bf k},\sigma}t_{\perp}({\bf k})(c_{{\bf
k},\sigma}^{\dagger (1)}c_{{\bf
k},\sigma}^{(2)}+{\rm h. c.})|n\rangle|^2}{E_n-E_0}.
\end{equation}
where $|n\rangle$ refers to an excited state of the uncoupled system.  
Since the crystal momentum is conserved,  the excited states must have the quantum number $\bf k $, which can be used to restrict the intermediate states. So we can write
\begin{equation}
\langle 0|{H_{\mathrm{eff}}}|0\rangle=\langle0|H_{1}+H_{2}|0 \rangle-\sum_{n\ne 0,{\bf k}}\frac{|\langle
0|\sum_{{\bf k'},\sigma}t_{\perp}({\bf k'})(c_{{\bf
k'},\sigma}^{\dagger (1)}c_{{\bf
k'},\sigma}^{(2)}+{\rm h. c.})|n{\bf k}\rangle|^2}{\varepsilon_{n,{\bf k}}(N+1)+\varepsilon_{n,{\bf k}}(N-1)},
\label{eq:secondorder}
\end{equation}
where $\varepsilon_{n,{\bf k}}(N\pm1)$ are the excitation energies for $(N\pm 1)$ particles, assumed to vanish near $\mathbf{k}a=(\pm \pi/2,\pm \pi/2)$. The tunneling matrix element in Eq.~\ref{eq:tmatrix1} is sharply peaked at ${\bf k}a=(\pi, 0)$ and symmetry related points; it is vanishingly small around the nodal regions. Moreover, if it vanishes at the nodes faster than the excitation energies, the sum can be restricted near the maxima. We denote such a region by the compact set$\{{\bf k}^{*}\}$. Then,
\begin{eqnarray}
\langle 0|{H_{\mathrm{eff}}}|0\rangle&\approx&\langle0|H_{1}+H_{2}|0 \rangle-\sum_{n\ne 0,{\bf k}\in \{{\bf k}^{*}\} }\frac{|\langle
0|\sum_{{\bf k'},\sigma}t_{\perp}({\bf k'})(c_{{\bf
k'},\sigma}^{\dagger (1)}c_{{\bf
k'},\sigma}^{(2)}+{\rm h. c.})|n{\bf k}\rangle|^2}{\varepsilon_{n,{\bf k}}(N+1)+\varepsilon_{n,{\bf k}}(N-1)}, \nonumber \\
&\approx&\langle0|H_{1}+H_{2}|0 \rangle-\frac{1}{2\langle\varepsilon_{n,{\bf k}}\rangle}\sum_{n\ne 0,{\bf k}\in \{{\bf k}^{*}\} }|\langle
0|\sum_{{\bf k'},\sigma}t_{\perp}({\bf k'})(c_{{\bf
k'},\sigma}^{\dagger (1)}c_{{\bf
k'},\sigma}^{(2)}+{\rm h. c.})|n{\bf k}\rangle|^2.
\end{eqnarray}
In the second line we have assumed that the excitation energy does not vanish in the set
$\{{\bf k}^{*}\}$ and can be pulled out of the sum by its average; the excitation energy of the $(N\pm 1)$ systems are the same modulo corrections of order $1/N$. We can now extend the summation region back to the entire Brillouin zone and add the terms $n=0$, because these represent vanishing contributions. Using the completeness of states, we can write the effective Hamiltonian as 
\begin{equation}
\langle 0|{H_{\mathrm{eff}}}|0\rangle \approx \langle0|H_{1}+H_{2}|0 \rangle-\frac{1}{D}\langle 0|\left[\sum_{{\bf k},\sigma}t_{\perp}({\bf k})(c_{{\bf
k},\sigma}^{\dagger (1)}c_{{\bf
k},\sigma}^{(2)}+{\rm h. c.})\right]^{2} |0\rangle ,
\end{equation}
where $D=2\langle\varepsilon_{n,{\bf k}}\rangle$. 

\begin{figure}[htb]
\begin{center}
\includegraphics[width=3.5in]{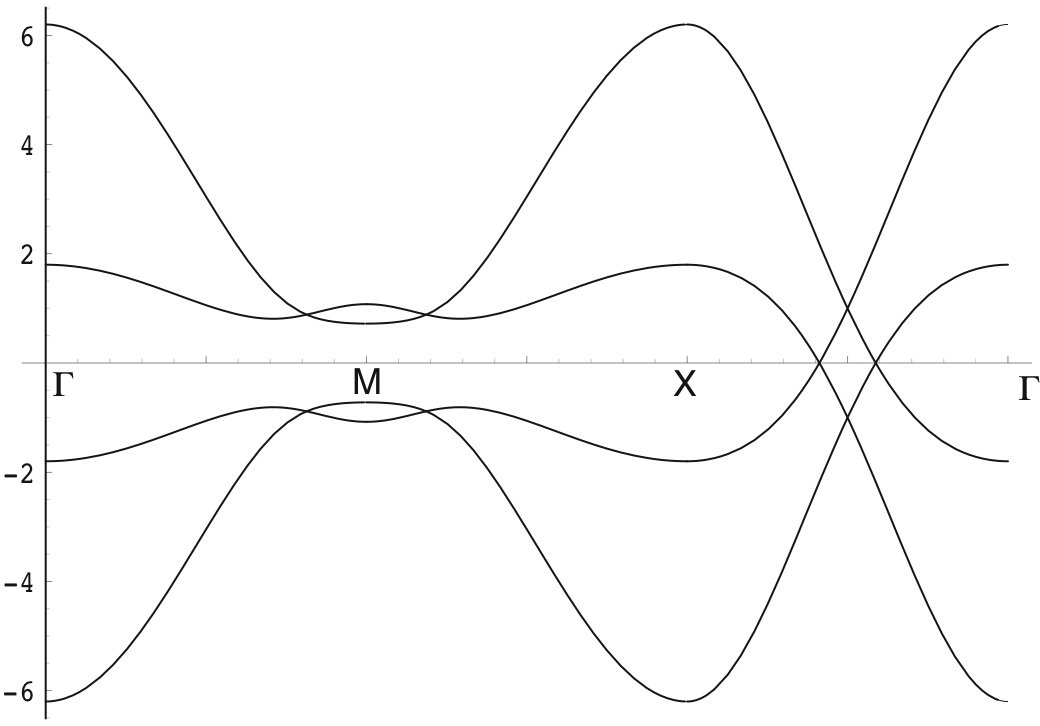} 
\end{center}
\caption[band]{\label{fig:band} The band structure of the coexisting DDW and DSC state. The energy is measured in units of $t$ and $t'=0.3$. The magnitudes of the order parameters are $|\phi|=0.4$ and $|\psi|=0.2$. The normal state chemical potential was chosen to be $\mu=-1$.}
\end{figure}

So far we have assumed that the translational symmetry of the ground state is preserved.  Now we exploit further robust symmetry arguments independent of the details of the microscopic physics. Consider the interesting case where the ground state breaks both translational and $U(1)$ symmetries but is a spin singlet. In particular, we consider the situation where a $d$-density wave (DDW) coexists but  competes with a $d$-wave superconductor (DSC).\cite{Chakravarty01} The important distinctions between DSC and DDW are useful to keep in mind. In the case of DSC, the pairing is between particles, and the symmetry of the orbital wave function determines the symmetry of the spin wave function. For DDW,  the pairing is between a particle and a hole, and the symmetry of the orbital wave function cannot enforce any symmetry requirement on the spin wave function; for example, there could be a singlet DDW as well as a triplet DDW.\cite{Nayak00} The respective order parameters for the singlet states are
\begin{eqnarray}
\langle c_{{\bf k},\sigma} c_{-{\bf k},\sigma'} \rangle&=&\frac{\psi}{2}(\cos k_{x}a - \cos k_{y}a)\delta_{\sigma,-\sigma'}: \mathrm{DSC},\\
\langle c_{{\bf k_+Q},\sigma}^{\dagger}c_{{\bf k},\sigma'} \rangle&=&i\frac{\phi}{2}(\cos k_{x}a - \cos k_{y}a)\delta_{\sigma,\sigma'}: \mathrm{DDW},
\end{eqnarray}
where ${\bf Q}=(\pi/a,\pi/a)$, $\psi$ is complex, but $\phi$ is a real scalar.  Note the Kronecker symbols involving the spin indices and the breaking of time reversal symmetry by the DDW order parameter, which is necessarily complex. 
For illustrative purposes, it is useful to plot the the band structure for the coexisting DDW and DSC states\cite{Chakravarty01}, as shown in Fig.~\ref{fig:band}. We have used a normal state energy dispersion $\varepsilon_{\mathbf{k}} = -2t (\cos{k_x a} + \cos{k_y a}) +4 t^{\prime} \cos{k_x a} \cos{k_y a}$. Note the energy gaps around the M point and the doubling of the nodes and the spectra  because of translational symmetry breaking. 

Noting that $t_{\perp}({\bf k})= t_{\perp}({\bf k + Q})$, we find that the effective Hamiltonian is
\begin{eqnarray}
H_{\mathrm{eff}}=H_{1}+H_{2}&-&{1\over
D}\sum_{{\bf k}\in \mathrm{RBZ}, \sigma}t_{\perp}({\bf k})^{2}(c_{{\bf k},\sigma}^{\dagger (1)}c_{-{\bf k},-\sigma}^{\dagger (1)}c_{-{\bf
k},-\sigma}^{(2)}c_{{\bf k},\sigma}^{(2)}+c_{{\bf k+Q},\sigma}^{\dagger (1)}c_{{\bf -k-Q},-\sigma}^{\dagger (1)}c_{{\bf
-k-Q},-\sigma}^{(2)}c_{{\bf k+Q},\sigma}^{(2)}\nonumber \\
&+&c_{{\bf k},\sigma}^{\dagger (1)}c_{{\bf
k},\sigma}^{(1)}c_{{\bf
k},\sigma}^{(2)}c_{{\bf k},\sigma}^{\dagger (2)}+c_{{\bf k+Q},\sigma}^{\dagger (1)}c_{{\bf
k},\sigma}^{(1)}c_{{\bf
k+Q},\sigma}^{(2)}c_{{\bf k},\sigma}^{\dagger (2)})+ {\rm h. c.},
\label{eq:Heff}
\end{eqnarray}
where RBZ stands for the reduced Brillouin zone, which is half the full Brillouin zone. In the simplest approximation all the coupling constants in Eq.~\ref{eq:Heff} are equal, but they need not be in a more refined derivation. This is obvious because the excitation energies are in principle different. Since we shall use this effective Hamiltonian merely to determine the form of the Ginzburg-Landau functional, this is of no consequence. It is now obvious that a competing order parameter, such as DDW, can reduce the enhancement resulting from the Josephson pair tunneling. This reduction is the analog of the cancellation found in the simplest BCS example discussed above. 

The important point in the derivation of the effective Hamiltonian is the structure of the tunneling Hamiltonian, which is sharply peaked in the region $\{{\bf k^{*}}\}$, away from the nodal region where it rapidly vanishes. The non-vanishing excitation energy leads to a controlled effective Hamiltonian via the one-step similarity transformation. In the absence of the special structure of the tunneling matrix element (constant in momentum, for example), other physical processes must be invoked to suppress the low energy hops. It may be possible to recover the same result from  orthogonality catastrophe in a non-Fermi liquid\cite{Chakravarty93} \footnote{The original interlayer theory\cite{Chakravarty93} was based on this scenario.} or from spin-charge separation, at least locally\cite{Khlebnikov96}. Similar arguments have been also presented in the stripe mechanism of spin gap proximity effect.\cite{Emery97}

For a  non-Fermi liquid, an electron creation operator of wavevector {\bf k} and spin $\uparrow$ acting
on the ground state  creates a linear superposition of states that carry the momentum $\bf k$ and spin $\uparrow$. 
The important point is that  the act of
inserting an electron into the many-body system cannot be renormalized away by defining a single
quasiparticle or a single superposition of a quasiparticle and a quasihole. If we like, this is the defining property of a non-Fermi liquid. This insertion
process must generate a superposition of multiparticle states, such that the total
momentum and spin  for each individual term in the superposition are still $\bf k$ and spin
$\uparrow$. But now the excitation energy is not uniquely related to $\bf k$. The states of low
energy are likely to contribute little to the sum because of the orthogonality catastrophe, which
generally leads to ground state overlaps of the form
\begin{equation}
\langle N| c_{{\bf k}\sigma}|N+1\rangle=
e^{-\alpha\int_{\omega_0({\bf k})}^{\omega_c}{d\omega \over \omega}}=\left({\omega_0({\bf k})\over
\omega_c}\right)^{\alpha} .\label{ortho}
\end{equation}
The quantity $\omega_0$ is a low energy cutoff, while $\omega_c$ is a high energy cutoff
of the order of the total bandwidth, and $\alpha$ is an orthogonality exponent that can depend on electron-electron interactions . Clearly the overlap vanishes as $\omega_0\to 0$, that is, in the nodal region. 
The above expression is simply the $z$-factor which must vanish at the Fermi surface in a non-Fermi
liquid state.  In the superconducting state, the states far above the gap are similar to the non-Fermi liquid  normal
state. However, in the regions of the Brillouin zone where the magnitude of the
gap is large, the orthogonality catastrophe will be cut off because $\omega_0$ in
Eq.~(\ref{ortho}) will be of the order of the energy gap $\Delta$, and the overlap factor will be $({\Delta\over \omega_c})^{\alpha}$. In the regions where the gap is vanishingly small, orthogonality
catastrophe will act with full force. 

\section{The Ginzburg-Landau energy functional}

\begin{figure}[tb]
\begin{center}
\includegraphics[width=3.5in]{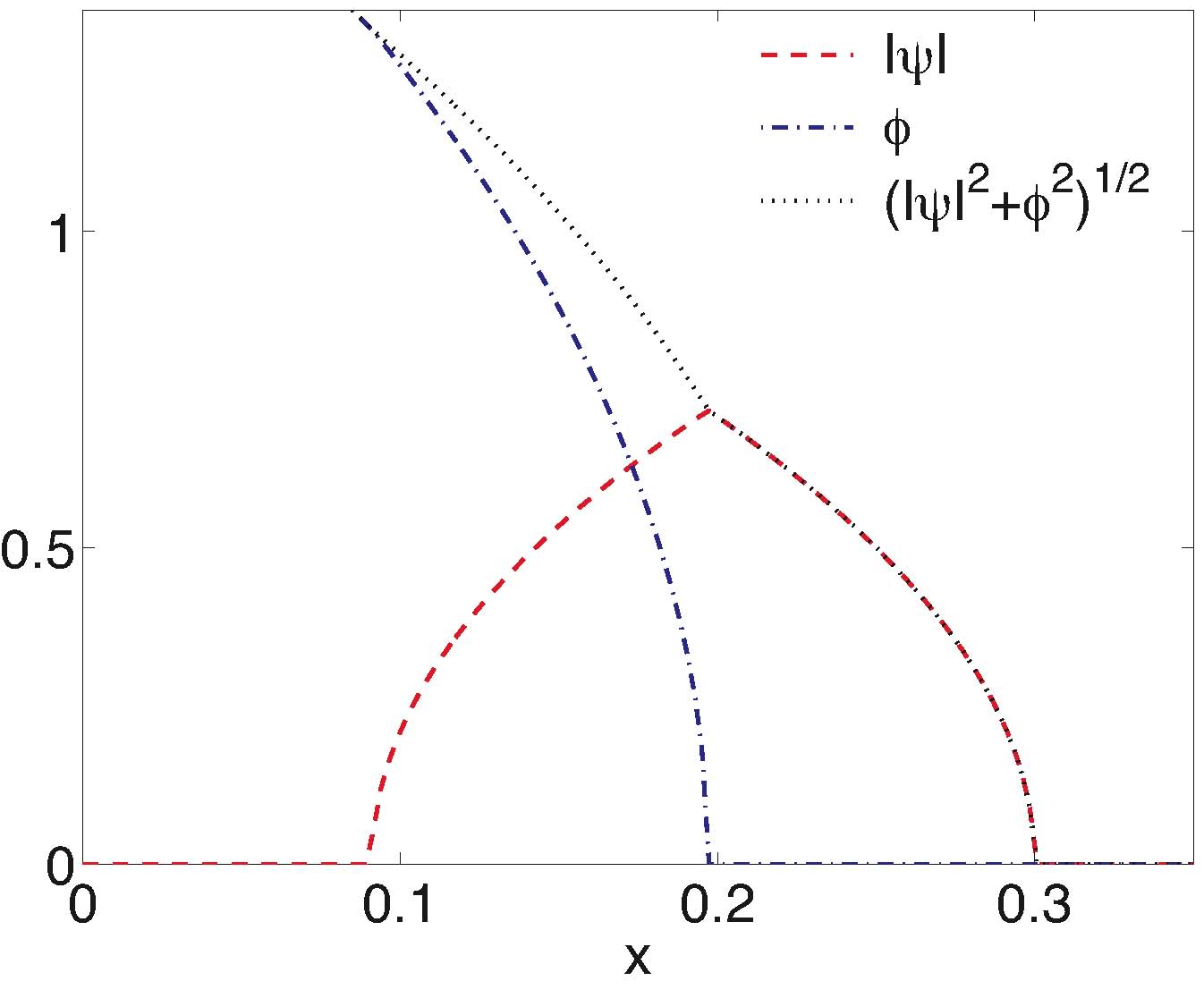} 
\end{center}
\caption[scdome]{\label{fig:scdome} $T=0$ phase diagram of a single-layer cuprate as a function of doping $x$.  $| \psi |$ is the magnitude of the superconducting order parameter and $\phi$ is the DDW order parameter.  We obtain a superconducting dome by taking $\alpha(x)=27 (x-0.22)$, $\alpha^{\prime}(x)=10(x-0.3)$, $\lambda=\lambda^{\prime}=1$, and $g=1.2$.}
\end{figure}

In this section we will try to capture the robust features of the phase diagram of multilayer cuprates using a suitable Ginzburg-Landau functional.  The advantage of Ginzburg-Landau theory is that it allows us to avoid the difficult problem of writing down an interacting Hamiltonian appropriate for the cuprates.  Instead, we can take our cue from the phenomenology of these materials and base our theory on the existence of specific broken symmetries.  Working at zero temperature simplifies matters further by limiting the number of adjustable parameters that appear in the theory.  If we assume that the ground state exhibits competing DDW and DSC orders, then the mean-field energy functional for an individual layer $j$ is given by
\begin{equation}
\mathcal{F}_{j}  = A \left[ \alpha^{\prime}
\left| \psi_{j}
\right|^{2} + \lambda^{\prime} \left| \psi_{j} \right|^{4} + \, \alpha \,
\phi_{j}^{2} + \lambda \, \phi_{j}^{4} + g \left| \psi_{j} \right|^{2}
\phi_{j}^{2} \, \right]
\end{equation}
where $\psi_j$ and $\phi_j$ are the superconducting and DDW order parameters, respectively, and $A$ is the area of the CuO$_2$ plane.  At $T=0$ the physics of these materials is (3+1)-dimensional, so we are justified in neglecting fluctuations.  To account for interlayer tunneling, we must add terms coupling layers $j$ and $j+1$; we will assume that longer-range coupling between layers is negligible.  From Eq.~\ref{eq:Heff} we see that the superconducting order parameters in neighboring layers will couple through a term of the form $-\rho_{c}(\psi_{j} \psi_{j+1}^{*} + \mbox{c.c.})$, where $\rho_c$ is a positive coupling constant---this is analogous to the pair-tunneling term of Section 2.  Ignoring (for the moment) interlayer DDW coupling, we arrive at the Ginzburg-Landau functional of Chakravarty {\em et al.} \cite{Chakravarty04}:
\begin{equation}
\mathcal{F}  =  {\displaystyle A \sum_{j=1}^{n}} \left[ \alpha^{\prime}
\left| \psi_{j}
\right|^{2} + \lambda^{\prime} \left| \psi_{j} \right|^{4}  + \, \alpha \,
\phi_{j}^{2} + \lambda \, \phi_{j}^{4} + g \left| \psi_{j} \right|^{2}
\phi_{j}^{2} - \rho_{c}
\left( \psi_{j} \psi_{j+1}^{*} + \mbox{c.c.} \right) \right]
\label{glnoJ}
\end{equation}
This is a zero-temperature energy functional for a single unit cell with $n$ CuO$_2$ planes.  Therefore the parameters $\alpha$, $\alpha^{\prime}$, $\lambda$, $\lambda^{\prime}$, $g$, and $\rho_c$ do not depend on $T$, although they can in principle depend on the index $j$.  Previously \cite{Chakravarty04} values to these parameters were assigned using two criteria:  (1) in the single-layer case ($n=1$, $\rho_{c}=0$), the phase diagram of Eq.~\ref{glnoJ} should produce a superconducting dome as a function of doping $x$; and (2) the transition temperature as a function of $n$ should describe an approximately bell-shaped curve peaked at $n=3$ (see Fig.~\ref{fig:Tc}).  As demonstrated in Fig.~\ref{fig:scdome}, the first criterion can be satisfied by setting $\lambda$, $\lambda^{\prime}$, and $g$ equal to constants and choosing $\alpha$ and $\alpha^{\prime}$ as linear functions of $x$.

\begin{figure}[tb]
\begin{minipage}[t]{6.5in}
\includegraphics[width=3in]{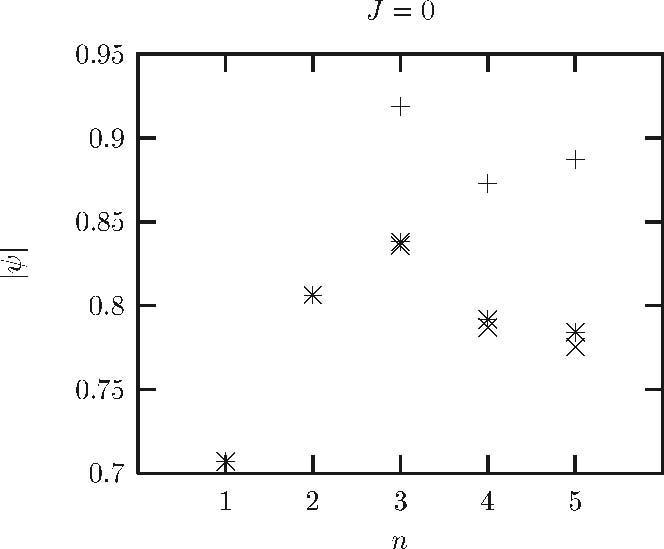} 
\hfill 
\includegraphics[width=3in]{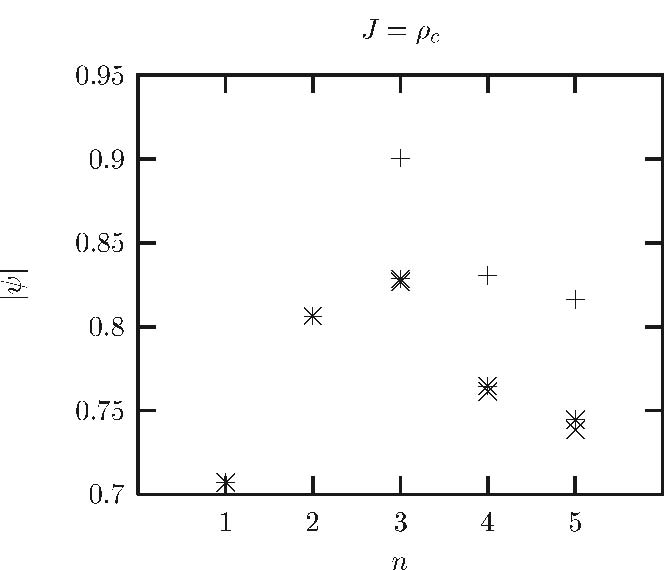}
\end{minipage}
\caption[noJ] { \label{fig:noJ} 
Left: $T=0$ superconducting order parameters of $n$-layer cuprates with no interlayer DDW coupling ($J=0$).  Doping, charge imbalance, and Ginzburg-Landau parameters are the same as in Chakravarty {\em et al.}\cite{Chakravarty04}  We plot three possible indicators of $T_c$:  $T_c \propto \psi_{\mbox{\scriptsize max}}$ ($+$), $T_c \propto \psi_{\mbox{\scriptsize avg}}$ ($\times$), and $T_c \propto \psi_{\mbox{\scriptsize rms}}$ ($\ast$).}
\caption[Jeq1]{\label{fig:Jeq1} Right: modification of the results in Fig.~\ref{fig:noJ} caused by interlayer DDW coupling $J=\rho_c$, where $\rho_c$ is the strength of the Josephson coupling.}
\end{figure}

To fulfill the second criterion, we must find a way to include both interlayer tunneling and the charge imbalance between the layers.  The latter occurs in multilayer samples with $n \geq 3$ because the outer layers feel a stronger electrostatic potential from the apical oxygens (Fig.~\ref{fig:multilayer}).  As a result, the inner layers of the unit cell become underdoped (with respect to the nominal doping $x$) and the outer layers become overdoped.  If the nominal doping is optimal, this causes a competing order parameter to appear on the inner layers, which ultimately decreases $T_c$.  Knight shift measurements performed by Kotegawa {\em et al.} \cite{Kotegawa01} provide values $\{x_{j}\}$ for optimally doped systems ($x \sim 0.2$) up to $n=5$.  Using these values, we can incorporate the charge imbalance into our functional through the parameters $\alpha=\alpha (x_{j})$ and $\alpha^{\prime}=\alpha^{\prime} (x_{j})$, whose doping dependence has already been fixed by the single-layer case.  Then the only adjustable parameter in Eq.~\ref{glnoJ} is the strength of the Josephson coupling, $\rho_c$, which can be chosen to optimize the shape of $T_c$ vs. $n$. Chakravarty {\em et al.}\cite{Chakravarty04} considered three possible indicators of $T_c$:   $\psi_{\mbox{\scriptsize max}} = \mbox{max} | \psi_j |$, $\psi_{\mbox{\scriptsize avg}} = \sum_{1}^{n} | \psi_j |/n$, and $\psi_{\mbox{\scriptsize rms}} = [ \sum_{1}^{n} | \psi_j |^2 / n ]^{1/2}$, where $\{ \psi_{j} \}$ are determined by a numerical minimization of Eq.~\ref{glnoJ} with respect to the $2n$ order parameters.  The calculation was performed with an open boundary condition, using the downhill simplex method with simulated annealing; the results, with $\rho_{c}=0.3$, are reproduced in Fig.~\ref{fig:noJ}.  The model clearly captures the peak of $T_c$ at $n=3$.

In the following sections we consider two modifications that can affect the shape of the curve for $n \geq 3$.  First, we examine the effects of interlayer DDW coupling, which were neglected in the original model, but should be present according to the effective Hamiltonian derived above.  Second, we look at what happens when the charge imbalance in the five-layer system is allowed to deviate from the numbers reported by Kotegawa {\em et al.}\cite{Kotegawa01}

\subsection{Interlayer DDW coupling}

The effective Hamiltonian derived in Section 3.2 suggests an interlayer DDW coupling of the form $J \phi_{j} \phi_{j+1}$, with $J>0$.  This is the DDW analog of the Josephson term:  DDW order parameters in neighboring layers prefer to align ``antiferromagnetically.''  (See the Appendix for a more detailed derivation of this result.)  We have therefore repeated the numerical minimization\cite{Chakravarty04} using the modified Ginzburg-Landau functional
\begin{equation}
\mathcal{F}^{\prime} = \mathcal{F} + A \sum_{j=1}^{n} J \, \phi_{j} \phi_{j+1}
\end{equation}
where $J$ is an adjustable (positive) parameter.  In Figs.~{\ref{fig:noJ} and \ref{fig:Jeq1}} we compare the results for $J=0$ to the results for $J=\rho_c$ ($\rho_c$ remains fixed at 0.3).  When the interlayer DDW coupling is as large as the Josephson coupling, the results for $n=4$ and $n=5$ change quite noticeably, but the peak at $n=3$ is robust.

\subsection{Charge imbalance}

The charge imbalance between inner and outer layers occurs only for $n \geq 3$, when the layers become crystallographically distinct (see Fig.~\ref{fig:multilayer}).  For $n=5$ it is also possible to distinguish between the innermost layer (IP$^{*}$) and its two neighbors (IP); however, Knight shift measurements \cite{Kotegawa01} suggest that the carrier concentration is roughly the same on all three.  The charge imbalance can therefore be quantified by the ratio $R= x_O / x_I$, where $x_O$ and $x_I$ denote the doping of the outer and inner planes, respectively.  If the nominal doping $x$ is known, then these quantities are also constrained by the relation $nx=2x_{O}+(n-2)x_{I}$.

\begin{figure}
\begin{minipage}[t]{6.5in}
\includegraphics[width=3in]{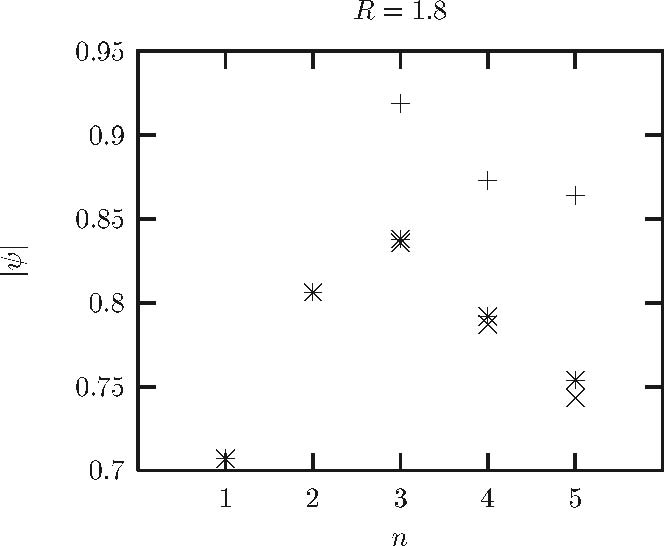} 
\hfill 
\includegraphics[width=3in]{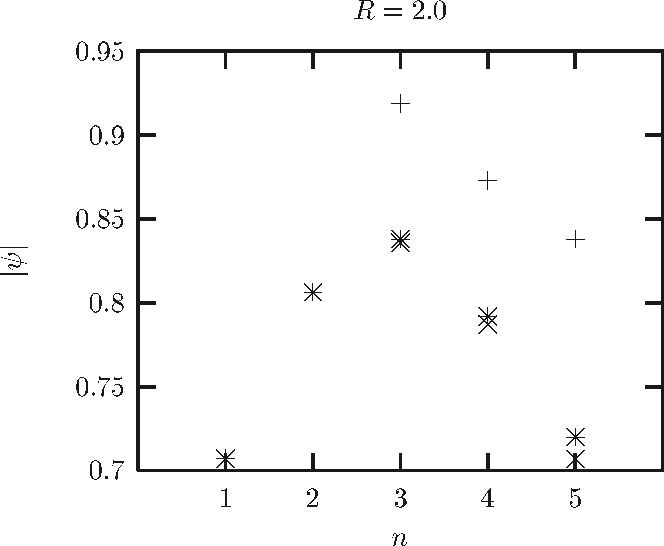}
\end{minipage}
\caption[r18]{\label{fig:r18} Left:  correction to the results of Fig.~\ref{fig:noJ} obtained with charge imbalance $R=1.8$ for $n=5$.  There is no interlayer DDW coupling ($J=0$).} 
\caption[r2]{\label{fig:r2} Right:  same as Fig.~\ref{fig:r18}, but with $R=2.0$ for $n=5$.  When the charge imbalance is this large, $\psi_{\mbox{\scriptsize avg}}$ and $\psi_{\mbox{\scriptsize rms}}$ return almost to the single-layer values at $n=5$, although $\psi_{\mbox{\scriptsize max}}$ remains considerably larger.}
\end{figure}

The results of the original calculation (Fig.~\ref{fig:noJ}) were obtained using values of $R$ and $x$ reported by  Kotegawa {\em et al.}\cite{Kotegawa01}:  $R=1.14$ for $n=3$, $R=1.49$ for $n=4$, and $R=1.64$ for $n=5$, with $x \sim 0.2$.  The numbers quoted for the three- and four-layer materials were averaged over values measured for several Hg- and Cu-based samples.  By contrast, the value of $R$ given for the five-layer system reflects a single measurement, on Cu1245.  A more recent experiment on the underdoped five-layer compound Hg1245 has yielded $R \sim 3.5$ for $x \sim 0.12$ \cite{Kotegawa04}.  Although a direct comparison of $R$-values for underdoped and optimally doped samples is probably misleading, there is at least reason to believe that the true value for an optimally doped five-layer material could deviate from 1.64.  Considering this possibility, we have repeated the minimization of Eq.~\ref{glnoJ} using somewhat larger values of $R$ for $n=5$.  The results are shown in Figs.~{\ref{fig:r18} and \ref{fig:r2}}.

\section{Final thoughts}

We have focused on the ground state and have pointed out that, despite competing effects, a systematic enhancement of superconductivity can be a universal feature over a limited range of the parameter space in the multilayer cuprates. And perhaps a better understanding can provide important directions of research. At finite temperatures, thermal fluctuations bring in a set of additional questions that we have not addressed. For example, phase fluctuations of the superconducting order parameter are expected to play an important role in underdoped cuprates,\cite{Emery95} where the superfluid density  is small and the coherence length is short. Indeed, coupling between the layers can suppress phase fluctuations and correspondingly increase $T_{c}$. This has been clearly demonstrated in model statistical mechanical systems.\cite{Carlson99}   It is an  interesting question to ask how the interplay between fluctuations and competing order will change the shape of the superconducting dome. In fact,  a part of the downturn of $T_c$ in the underdoped regime can also  be due to  phase fluctuations rather than a competing order parameter. To add more complexity, it may be necessary to step beyond an order parameter theory and assess the role of excitations of nodal quasiparticles in the degradation of $\mathrm{T_{c}}$. The distinguishing feature of the mechanism of the superconducting dome discussed here is unique to high-$\mathrm{T_{c}}$  cuprate superconductors, however.

We are not entirely sanguine that the data shown in Fig.~\ref{fig:Tc} is the final answer, and it would be important to explore further the dependence of $\mathrm{T_{c}}$  on the number of layers $n$. From our theoretical perspective, it is natural to suggest that with increasing $n$, $\mathrm{T_{c}}$  should saturate to the single layer value and the enhancement is only an intermediate scale effect. As we have discussed, with increasing $n$, the outer planes get overdoped, thereby decreasing $T_{c}$, and the inner planes get underdoped, which also reduces $\mathrm{T_{c}}$ , as long as we believe in the existence of a superconducting dome in a single-plane material.  This implies that there must be an intermediate layer for which the doping is close to the nominal optimal value, which determines $\mathrm{T_{c}}$ . It is difficult to rigorously argue that  this should be the same as the single-plane $\mathrm{T_{c}}$, but we suspect that this is the case. In other words, competing order ultimately results in an exact cancellation of the enhancement. Of course, other physical processes may also arise. It would be extraordinarily interesting to establish the correct $\mathrm{T_{c}}$ versus $n$ behavior from careful future experiments.

Recent experiments suggest that the competing order that is nucleated in the inner layers is antiferromagnetic.\cite{Kotegawa04} We are not entirely sure that this the case. Impurity effects can often result in spurious antiferromagnetic order.\cite{Sidis01} There are further reasons for believing that the aniferromagnetic order parameter may not be the key competing order parameter from the ``no-mixing of antiferromagnetism and $d$-wave superconductivity'' experiments of Bozovic {\em et al.}\cite{Bozovic03} However, $\mu$SR (muon spin rotation) measurements in multilayer cuprates\cite{Tokiwa03} find strikingly strong evidence of antiferromagnetism in the inner planes, which appear, however, to be difficult to understand. We suspect the positive muons to sit close to the apical oxygens and thus should result in a signal dominated by the closest outer planes rather  than the inner planes. It is therefore  unclear why the  $\mu$SR signals reflect magnetic moments with magnitudes almost as large as the undoped antiferromagnet. Clearly, it is desirable to explore these questions using many different precise experimental tools, as the enhancement of the superconducting transition temperature is one of the most important questions in condensed matter physics. From the theoretical perspective, however, there is little doubt that a charge imbalance sets in in multilayer cuprates, because this effect is the result of robust energetics involving merely electrostatics.\cite{Distasio90,Haines92}

\paragraph{Note added:} W. Zwerger has drawn our attention to a paper by R. A. Ferrell, {\em Phys. Rev. B} {\bf 38}, 4984-4985 (1988), where it is shown that the root of the exact cancellation of single particle tunneling and pair tunneling discussed in Sec. 2 is ``Anderson's dirty superconductor theorem.'' Conversely, this cancellation will fail when the assumptions of this theorem fail. The simplest such failure is when the gaps of the two superconductors are unequal (proximity effect). Two other recent papers on interlayer tunneling are,  J. -B. Wu, M. -X Pei, and Q. -H. Wang, {\em Phys. Rev. B} {\bf 71}, 172507/1-4 (2005) and M. Mori and S. Maekawa, {\em Phys. Rev. Lett.} {\bf 94}, 137003/1-4 (2005). We thank the authors for pointing these out to us.

\appendix
\section{Antiferromagnetic DDW coupling}

It is sufficient to consider two layers in a pure DDW (rather than coexisiting DDW+DSC) phase.  Suppose that $H_1$ and $H_2$ both have the form of a mean-field DDW Hamiltonian \cite{Chakravarty01} on a square lattice, given by
\begin{equation}
H_{\mbox{\scriptsize DDW}}  =  \sum_{\mathbf{k} \in \mbox{\scriptsize RBZ}, \sigma} \left[ ( \varepsilon_{1 \mathbf{k}} + \varepsilon_{2 \mathbf{k}} - \mu )  c_{\mathbf{k} \sigma}^{\dagger} c_{\mathbf{k} \sigma}
+ ( -\varepsilon_{1 \mathbf{k}} + \varepsilon_{2 \mathbf{k}} - \mu )  c_{\mathbf{k}+\mathbf{Q}, \sigma}^{\dagger} c_{\mathbf{k}+\mathbf{Q}, \sigma}+ (i \, W_{\mathbf{k}} c_{\mathbf{k} \sigma}^{\dagger} c_{\mathbf{k}+\mathbf{Q},\sigma} + \mbox{h.c.}) \right]
\label{hddw}
\end{equation}
We assume a simple $t-t^{\prime}$ band structure, with $\varepsilon_{1 \mathbf{k}} = -2t (\cos{k_x a} + \cos{k_y a})$ and $\varepsilon_{2 \mathbf{k}} = 4 t^{\prime} \cos{k_x a} \cos{k_y a}$.  The DDW order parameter $\phi$ enters through $W_{\mathbf{k}}=(W_{0}/2)(\cos{k_{x}a}-\cos{k_{y}a})$, where $W_0$ is proportional to a microscopic energy scale times $\phi$.  Employing a Bogoliubov transformation, we can diagonalize Eq.~\ref{hddw} to obtain the simple two-band Hamiltonian
\begin{equation}
H_{\mbox{\scriptsize DDW}} = \sum_{\mathbf{k} \in \mbox{\scriptsize RBZ}, \sigma} \left[ (E_{\mathbf{k}}^{+}-\mu ) \alpha_{\mathbf{k} \sigma}^{\dagger} \alpha_{\mathbf{k} \sigma} + (E_{\mathbf{k}}^{-}-\mu ) \beta_{\mathbf{k} \sigma}^{\dagger} \beta_{\mathbf{k} \sigma} \right]
\end{equation}
The fermion operators $\alpha_{\mathbf{k} \sigma} = u_{\mathbf{k}} \, c_{\mathbf{k} \sigma} + iv_{\mathbf{k}} \, c_{\mathbf{k}+\mathbf{Q}, \sigma}$ and $\beta_{\mathbf{k} \sigma} = \mbox{sgn}(W_{\mathbf{k}}) ( v_{\mathbf{k}} \, c_{\mathbf{k} \sigma} -i u_{\mathbf{k}} \, c_{\mathbf{k}+\mathbf{Q}, \sigma} )$ destroy quasiparticles in the energy bands $E_{\mathbf{k}}^{\pm} = \varepsilon_{2 \mathbf{k}}\pm ( \varepsilon_{1 \mathbf{k}}^2 + W_{\mathbf{k}}^2 )^{1/2}$.  The complex coherence factors result from broken time-reversal symmetry; their real parts are given by
\begin{equation}
u_\mathbf{k} = {W_\mathbf{k} \over \left( W_\mathbf{k}^2 +  \left[ (\varepsilon_{1 \mathbf{k}}^2 + W_\mathbf{k}^2)^{1/2} - \varepsilon_{1 \mathbf{k}} \right]^2 \right)^{1/2}} \; , \;
v_\mathbf{k} = {{ ( \varepsilon_{1 \mathbf{k}}^2 + W_\mathbf{k}^2 )^{1/2} - \varepsilon_{1 \mathbf{k}}} \over \left( W_\mathbf{k}^2 +  \left[ (\varepsilon_{1 \mathbf{k}}^2 + W_\mathbf{k}^2)^{1/2} - \varepsilon_{1 \mathbf{k}} \right]^2 \right)^{1/2}}
\end{equation}
Note that $u_\mathbf{k}$ depends on the sign of $W_\mathbf{k}$ and therefore on the sign of $\phi$---this is unimportant for an isolated layer, but very important for two coupled layers, as we will demonstrate below.  Equally important are the symmetries of the coherence factors under the transformation $\mathbf{k} \rightarrow \mathbf{k} + \mathbf{Q}$:  $u_{\mathbf{k}+\mathbf{Q}} = - \mbox{sgn}(W_\mathbf{k}) v_\mathbf{k}$ and $v_{\mathbf{k}+\mathbf{Q}} = \mbox{sgn}(W_\mathbf{k}) u_\mathbf{k}$, which means that the eigenstates created by $\alpha_{\mathbf{k}+\mathbf{Q}, \sigma}^{\dagger}$ and $\beta_{\mathbf{k}+\mathbf{Q}, \sigma}^{\dagger}$ are the same as those with momentum $\mathbf{k}$, up to a phase factor.  In terms of these quasiparticle operators, the zero-temperature DDW ground state $| 0 \rangle$ can be defined for arbitrary filling fraction by $\alpha_{\mathbf{k} \sigma}^{\dagger} | 0 \rangle \propto \Theta ( E_\mathbf{k}^+ -\mu)$ and $\beta_{\mathbf{k} \sigma}^{\dagger} | 0 \rangle \propto \Theta ( E_\mathbf{k}^- - \mu)$.

Now consider the unperturbed ground state manifold $|0 \rangle = |0 \rangle_{(1)} \otimes |0 \rangle_{(2)}$ of two identical layers ($|\phi_{1}|=|\phi_{2}|$).  There are two degenerate ground states:  $|0 \rangle=|F\rangle$, in which $\phi_{1}=\phi_{2}$, and $|0\rangle = |AF\rangle$, in which $\phi_{1}=-\phi_{2}$.  Since the DDW excitation spectrum is gapped near the $(\pi,0)$ points, the arguments of Section 3.2 still apply.  The effective Hamiltonian for the low energy sector is
\begin{equation}
H_{\mathrm{eff}}=H_{1}+H_{2}-\frac{H_{T}^{2}}{D}
\end{equation}
Generally we expect the degeneracy between $|F\rangle$ and $|AF\rangle$ to be lifted by the third term, so that the true ground state is some linear combination of the two.  What we need to compute, then, are the matrix elements of $H_{T}^{2}$ in this subspace.  Exploiting the symmetries of the coherence factors, along with the additional symmetry $t_{\bot}(\mathbf{k}+\mathbf{Q}) = t_{\bot} (\mathbf{k})$, it is straightforward to show that
\begin{equation}
H_{T} | 0 \rangle  =  \sum_{\mathbf{k} \in \mbox{\scriptsize RBZ}, \sigma} \left\{ t_{\bot} (\mathbf{k}) \, \mbox{sgn} \left[ W_\mathbf{k}^{(2)} \right] \left[ u_\mathbf{k}^{(1)} v_\mathbf{k}^{(2)}  - \, v_\mathbf{k}^{(1)} u_\mathbf{k}^{(2)} \right] \alpha_{\mathbf{k} \sigma}^{\dagger (1)} \beta_{\mathbf{k} \sigma}^{(2)} | 0 \rangle 
+  \left[ 1 \rightarrow 2, 2 \rightarrow 1 \right] \right\},
\end{equation}
provided the system is at or below half-filling.  For $| 0 \rangle = | \mbox{F} \rangle$, we have $v_\mathbf{k}^{(1)} = v_\mathbf{k}^{(2)}$ and $u_\mathbf{k}^{(1)} = u_\mathbf{k}^{(2)}$, which obviously gives $H_{T} | \mbox{F} \rangle = 0$.  This cancellation of coherence factors does not occur when $| 0 \rangle = | \mbox{AF} \rangle$, because then $v_\mathbf{k}^{(1)} = v_\mathbf{k}^{(2)}$ and $u_\mathbf{k}^{(1)} = -u_\mathbf{k}^{(2)}$.  Thus we have arrived at a rather unexpected result:  the two states are not mixed at {\em any} order in perturbation theory because $H_{T} |F\rangle$ vanishes identically.  Instead, the state $|F\rangle$ remains at the same energy while the state $|AF\rangle$ has its energy lowered by an amount
\begin{equation}
\Delta E_{0}^{AF} = -\frac{16}{D} \sum_{\mathbf{k}\in\mathrm{RBZ}} t_{\bot}(\mathbf{k})^{2}u_{\mathbf{k}}^{2}v_{\mathbf{k}}^{2} \, \Theta (E_{\mathbf{k}}^+ -\mu ) \Theta (\mu-E_{\mathbf{k}}^- )
\end{equation}
Here we have dropped the superscripts since $u_{\mathbf{k}}^{2}$ and $v_{\mathbf{k}}^{2}$ are the same in the two layers.  The energy scale $D$ is set by some combination of the DDW gap $W_{0}$ and the bandwidth $t$.

\acknowledgments  
We thank S. Kivelson, D. J. Scalapino and R. S. Thompson  for discussions. This work was supported by a grant from the National Science Foundation: NSF-DMR 0411931.

\bibliography{ILT}   
\bibliographystyle{spiebib}   

\end{document}